\documentclass[twocolumn]{aastex701}
\usepackage{array}
\usepackage{amsmath}
\usepackage{datetime}
\usepackage{diagbox}
\usepackage{enumerate}
\usepackage{esint}
\usepackage{longtable}
\usepackage{extarrows}
\usepackage{bm}
\usepackage{graphicx}
\usepackage{lineno}
\linenumbers

\newcommand{\be}{\begin{equation}}
\newcommand{\ee}{\end{equation}}

\begin{document}

\title{Impact of Resonant Compton Scattering on Magnetar X-Ray Polarization with QED Vacuum Resonance}

\author[orcid=0009-0002-1484-9278]{Tu Guo}
\affiliation{School of Physics and Astronomy, Shanghai Jiao Tong University, Shanghai 200240, China}
\affiliation{Tsung-Dao Lee Institute, Shanghai Jiao Tong University, Shanghai, 201210, China}
\email[]{2743885833@sjtu.edu.cn}

\author[orcid=0000-0002-1934-6250]{Dong Lai} 
\affiliation{Tsung-Dao Lee Institute, Shanghai Jiao Tong University, Shanghai, 201210, China}
\affiliation{Center for Astrophysics and Planetary Science, Department of Astronomy, Cornell University, Ithaca, NY 14853, USA}
\email[show]{donglai@sjtu.edu.cn}

\begin{abstract}
Recent observations have revealed significant soft X-ray polarization from several quiescent magnetars, including the intriguing $90^\circ$ polarization angle (PA) swing as a function of photon energy for some sources. We thus present a general semi-analytical framework for calculating energy-dependent soft X-ray polarization signatures from magnetars, consistently incorporating both QED vacuum resonance in the atmosphere and resonant Compton scattering (RCS) in the magnetosphere. Starting from the polarized radiative transfer equation for RCS and treating vacuum-resonance-induced mode conversion as an input, we employ a first-order approximation in RCS optical depth to evaluate the effects of magnetospheric plasma density (which depends on magnetic twist), drift velocity, temperature, and viewing geometry on the observed radiation. Our analysis reveals that magnetic twist and plasma drift velocity are the critical parameters controlling the impact of RCS on both the absolute polarization degree and its variation across the soft X-ray spectrum. We find that sufficiently strong RCS can wash out the PA swing caused by vacuum resonance. Furthermore, in addition to the QED vacuum resonance effect, significant relativistic signatures arising from plasma drift velocity ($\beta_0 \gtrsim 0.5$) may introduce an extra $90^\circ$ PA swing in the spectrum. Our calculation framework, based on the single-scattering approximation, bypasses the need for complex, {multidimensional} Monte Carlo simulations, providing an analytical pathway for modeling full-surface emission and rotational-phase-resolved radiation from magnetic neutron stars {in} support of current and future X-ray polarization missions.
\end{abstract}



\section{Introduction}

Magnetars are isolated neutron stars endowed with extremely strong magnetic fields \citep{D&T1992,T&D1995,K1998natur,Kaspi2017}. Their surface dipole magnetic fields inferred from spin-down measurements typically {reach} $B_{\rm dip} \sim 10^{14}$--$10^{15}~\mathrm{G}$. Historically, the magnetar subclasses of Soft Gamma Repeaters (SGRs) and Anomalous X-ray Pulsars (AXPs) were distinguished by their burst behaviors, but both emit persistent soft X-rays ($0.5$--$10~\mathrm{keV}$) in quiescence with typical luminosities of $L_{\rm X} \sim 10^{33}$--$10^{36}~\mathrm{erg\ s^{-1}}$, much larger than {their spin-down} luminosities. Their quiescent X-ray spectra are commonly described by {a} combination of a thermal blackbody with $kT \sim 0.5$--$1~\mathrm{keV}$ and a non-thermal power law with photon index $\Gamma \sim 2$--${4}$ \citep{Woods&Thompson2006}. They also emit hard X-rays extending to $\sim 100~\mathrm{keV}$ {with no evidence of a spectral break below 70-100 keV}, indicating {an} active corona surrounding magnetars \citep{Kaspi2017}. Despite the rich phenomenology, the integrated X-ray spectrum alone is often inadequate to disentangle the degeneracy among different emission mechanisms and magnetospheric physics. {However,} X-ray polarization measurements can help break this degeneracy and provide constraints on various physical parameters and magnetospheric structure \citep{Fernandez2011,IXPE2022,Taverna2024}.

In recent years, the Imaging X-ray Polarimetry Explorer (IXPE) mission \citep{IXPE2016} has enabled polarimetry measurements in the $2$--$8~\mathrm{keV}$ band, delivering high-significance detections across multiple sources, including several magnetars. The IXPE observations of magnetars have revealed {diverse energy-dependent polarization behaviors} in {the} soft X-ray band: \added{some} \citep[e.g., 4U 0142+61 in][]{Sci2022} exhibit a sharp $\sim 90^\circ$ polarization angle (PA) swing at around ${4-5~\mathrm{keV}}$, \added{some might show signs of two $\sim 90^\circ$ swings (e.g., SGR 1806-20 in \cite{Turolla2023}, although with low statistical significance),} while others \citep[e.g., 1RXS J170849.0-400910 in][]{Zane2023} maintain a {nearly energy-independent constant} polarization angle.

To explain these spectral polarization features {from IXPE results}, current theoretical efforts mainly focus on two physical mechanisms: emission from {a} condensed surface \citep{Sci2022,Taverna2024}, followed by resonant Compton {scattering} (RCS) in the magnetosphere \citep{Fernandez2011,Taverna2014}, and the mode conversion effect via vacuum resonance in the atmospheric plasma under strong magnetic fields \citep{Lai&Ho2002,Lai&Ho2003a,Lai2023,Ruth2024}. A condensed surface requires sufficiently low temperature and high surface binding energy, which {remain} uncertain \citep{Medin&Lai2006a,Medin&Lai2006b,Medin&Lai2007,Potekhin&Chabrier2013}. RCS by charged particles in a twisted magnetosphere can modify the emergent radiation spectrum and polarization profile. This process has been modeled in both semi-analytical 1D frameworks \citep{L&G2006} and {multidimensional} Monte Carlo simulations with detailed spectral fits \citep{Fernandez2007,Nobili2008a,Rea2008,Zane2009}, and further extended to polarization predictions \citep{Fernandez2011,Taverna2014,Taverna2020,Taverna2024}.

Vacuum resonance, on the other hand, arises from the competition between plasma-induced birefringence and vacuum birefringence due to quantum electrodynamics (QED) in strong magnetic fields. When a photon {crosses} the resonant layer, it may undergo adiabatic mode conversion, leading to a significant change in the polarization \citep{Lai&Ho2002,Lai&Ho2003a,Lai&Ho2003b}. After leaving the atmosphere, the photon may experience significant RCS in the extended magnetosphere or corona, leading to further modifications in polarization.

In this work, we present a general three-dimensional semi-analytical framework that unifies the treatment of vacuum resonance in the atmosphere and RCS in the magnetosphere, with the goal of obtaining the soft X-ray polarization of {magnetic} neutron stars. To preserve the physical transparency and geometric modularity of our calculation without relying on full Monte Carlo simulations of radiative transport, we use {the} single-scattering approximation. To focus on the key physics, we adopt simplified yet testable idealized assumptions for the polarized surface emission, {including} the vacuum resonance effect. The structure of this paper is as follows: Section~\ref{sec:model} introduces the physical model and computational procedure; Section~\ref{sec:validation} presents the validation of our fiducial model; Section~\ref{sec:results} investigates the X-ray polarization signals for various physical parameters; Section~\ref{sec:conclusion} provides the summary and outlook.

\section{Physical Framework} \label{sec:model}

To model how resonant Compton scattering and QED vacuum resonance jointly shape the X-ray polarization from magnetars, we distinguish the main physical processes with their characteristic locations. This section presents our physical framework for {the} global magnetosphere, polarized emission and propagation, RCS formalism, and computational workflow.

\subsection{Global Magnetosphere Model}

One of the key characteristics of magnetars is the presence of a highly twisted and dynamically evolving toroidal magnetic field in their magnetospheres. This arises from the magnetic stress built up in the crust, which induces instabilities, leading to displacement of surface elements and distortion of the magnetic field lines anchored on them. {\citet{TLK}} present a general self-similar, force-free, globally twisted magnetic field model that has been widely used in previous studies \citep[e.g.,][]{Fernandez2007,Nobili2008a,Fernandez2011}.

For the purposes of this work, we adopt a simple parametrized model in which a toroidal component proportional to the poloidal field is added, i.e.,
\begin{equation} \label{ma}
\boldsymbol{B}(\boldsymbol{r})=\frac{B_\mathrm{p}}{2}\left( \frac{r}{R} \right)^{-3} \left( 2\cos{\theta}, \sin{\theta},\xi_\tau \sin{\theta} \right),
\end{equation}
where $B_{\mathrm p}$ is the surface polar dipole field strength, $R$ is the magnetar radius, and the parameter $\xi_{\tau}$ quantifies the global twist (hence the current distribution). This choice is not strictly force-free, but it captures a quasi-static configuration adequate for our semi-analytic treatment.

We assume {that} the magnetospheric plasma consists of {a single charge species} and that the ions are stationary. Given $\boldsymbol B$, the spatial distribution of current-carrying electrons along the closed field lines {measured in the stellar frame} is prescribed as \citep{TLK,Nobili2008a,Taverna2020}
\begin{equation}\label{density}
n_e(\boldsymbol r,\beta_0)=\frac{(\xi_\tau \beta_0^{-1})\ B(\boldsymbol{r})}{4\pi e\ r},
\end{equation}
where $B=|\boldsymbol{B}(\boldsymbol{r})|$. {We note that kinetic modeling of external currents in magnetars suggests that the real magnetospheric plasma may be a denser electron--positron pair plasma with high $e^\pm$ multiplicity \citep[][see Section~\ref{sec:conclusion} for further discussion]{Thompson2008,Beloborodov2013,Thompson2020,Zhang2025}}. We denote {by} $\beta_e\in(-1,1)$ the electron velocity (normalized by $c$) along the field lines, and $\beta_0 = |\langle \beta_e \rangle|$ is the mean drift speed. To determine the full phase-space distribution, we assume a one-dimensional, relativistic, thermal (Maxwell--J\"uttner) velocity distribution along the field line, boosted by $\beta_{0}$ \citep[see also][]{Nobili2008a,Nobili2008b}:
\begin{equation}
f_1(\beta_e)\,\mathrm{d}\beta_{e}=\frac{\gamma_0 \gamma_e^3 (1 - \beta_e\beta_0)\exp{(-\gamma'/\Theta_e)}}{2K_1(1/\Theta_{e})}\,\mathrm{d}\beta_{e},
\end{equation}
with $\Theta_{e} = {kT_{e}}/{m_{e}c^2}$, $\gamma' = \gamma_{e}\gamma_{0}(1 - \beta_{e}\beta_{0})$, and $K_{1}$ the modified Bessel function. This boosting captures the net drift from the magnetic north to magnetic south. We assume {that} the magnetospheric temperature $T_{e}$ is uniform, and treat $T_{e}$ and $\beta_{0}$ as free parameters. Overall, our magnetosphere model is characterized by three parameters: $(\xi_{\tau}, \beta_{0}, T_{e})$.

\subsection{Seed Polarized Radiation \& Propagation}

\subsubsection{Surface Emission Model} \label{sec:surfaceemission}

The bulk of the ${2}$--${8~\mathrm{keV}}$ photons probed by IXPE originate from the magnetar surface. To model the spectrum and polarization of the surface emission, it is necessary to construct self-consistent magnetic NS atmosphere models that take QED effects into account \citep[e.g.,][]{Ho&Lai2003,Lai&Ho2003a,vanAdelsberg2006,Shabaltas&Lai2012,Taverna2020,Caiazzo2022}. Here we summarize the
key features of this emission \citep[see also][]{Lai2023}.

X-ray photons (with $E=\hbar \omega \ll E_B=\hbar\omega_{{B}} = \hbar e B / m_e c=1158 B_{14}~{\mathrm{keV}}$) propagating in the magnetized NS atmospheric plasma have two polarization modes: the ordinary mode (O-mode) is mostly polarized parallel to the $\boldsymbol{k}-\boldsymbol{B}$ plane, while the extraordinary mode (X-mode) is perpendicular to it, where $\boldsymbol{k}$ is the photon wave vector and $\boldsymbol{B}$ is the external magnetic field. The two modes have very different absorption and scattering opacities, with $\kappa_{\mathrm{X-mode}}\sim u_e^{-1}\kappa_{\rm O-mode}$, where $u_e\equiv (\omega_B/\omega)^2\gg 1$. Thus, the two modes have very different photosphere densities, with $\rho_{\rm X-mode}\gg\rho_{\rm O-mode}$, and the emergent radiation from the NS surface is dominated by the X-mode.

However, the effect of vacuum polarization can change the above
picture significantly. In particular, the ``competition'' between QED-induced vacuum birefringence and plasma-induced birefringence gives rise to a vacuum resonance; for a photon of energy $E$, this occurs at the density
\be
\rho_V\simeq 0.964\,Y_e^{-1}B_{14}^2E_1^2 f^{-2}~{\mathrm{g~cm}^{-3}},
\label{eq:densvp}
\ee
where $Y_e=\langle Z/A\rangle$ is the electron fraction,
$E_1=E/(1~{\mathrm{keV}})$, and $f=f(B)$ is a slowly varying function of
$B$ and is of order unity.

For $\rho\gtrsim \rho_V$ (where the plasma
effect dominates the dielectric tensor) and $\rho\lesssim\rho_V$ (where vacuum polarization dominates), the photon modes (for typical
$\theta_{\rm kB}\neq 0$) are almost linearly polarized, i.e.{,} the linear X-mode and O-mode; at $\rho=\rho_V$, however, the normal modes become circularly polarized. The importance of this vacuum resonance lies in the {mode-conversion} phenomenon \citep{Lai&Ho2002,Lai&Ho2003a}: {an} X-mode photon emerging from its photosphere (at $\rho_{\rm X-mode}$) and propagating {outward} can convert to the O-mode at $\rho_V$, provided that its energy satisfies
\be
E\gtrsim E_{\rm ad}=2.52\,\bigl(f\,\tan\theta_{\rm kB} \bigr)^{2/3}
\!\left(\frac{1{~\mathrm{cm}}}{{H_\rho}}\right)^{1/3}\!{\mathrm{keV}},
\label{condition}\ee
where $H_\rho=|\mathrm{d}s/\mathrm{d}\ln\rho|$ is the density scale
height (evaluated at $\rho=\rho_V$) along the ray. A low-energy ($E\lesssim E_{\rm ad}$) photon would not experience this conversion. In typical magnetar atmospheres, $\rho_{\rm X-mode}\gg \rho_V$ is always satisfied. Depending on the relative locations of the O-mode photosphere and vacuum resonance, the X-ray polarization signatures are qualitatively different. Assuming a fully ionized atmosphere (with ion {charge} $Ze$ and mass $Am_p$), {\citet{Lai2023}} finds the ratio
\be
\rho_V/\rho_{\rm O-mode}=(B/B_{\rm OV})^2,
\ee
with
\be
B_{\rm OV}=7.8\!\times\! 10^{13}\left(\!\frac{{\mu} g_2{\cos\alpha}}{Z{\cal G}E_1{\sin^2\!
\theta_{\rm kB}}}\!\right)^{\!1/4}\!\!\left(\!\frac{f}{{T_6^{1/8}}}\!\right)
{\mathrm{G}},\label{eq:Bov}
\ee
where $\mu=A/(1+Z)$ is the ``molecular'' weight, $g_2$ is the surface gravity in units of $2\times 10^{14}~{\mathrm{cm~s^{-2}}}$, $T_6$ is {the} surface temperature in units of $10^6~{\mathrm{K}}$, and ${\cal G}=1-\exp(-E/kT)$. For magnetars with surface magnetic fields satisfying $B\lesssim B_{\rm OV}$, the vacuum resonance lies at a lower density than $\rho_{\rm O-mode}${;} the emergent radiation is dominated by the X-mode for $E\lesssim E_{\rm ad}$ and by the O-mode for $E\gtrsim E_{\rm ad}$, i.e., there is a $90^\circ$ rotation of the polarization angle between $E\lesssim E_{\rm ad}$ and $E\gtrsim E_{\rm ad}$. On the other hand, for $B\gtrsim B_{\rm OV}$, the vacuum resonance lies between $\rho_{\rm O-mode}$ and $\rho_{\rm X-mode}$, and the emergent radiation from the surface is dominated by {the} X-mode for all {values of} $E$.

Quantitative calculations of the atmospheric emission (spectrum and polarization) are fully described in previous works \citep[e.g.,][]{Ho&Lai2003,vanAdelsberg2006,Lai2023}. In this paper, since our focus is on the effect of magnetospheric {scattering}, we adopt a simple prescription for the surface emission. For a given spot (located at the polar angles $\theta_s,\phi_s$) with temperature $T_s$, the total radiative intensity (in the stellar frame) is given by
\be
I_\omega^{{\mathrm{(tot)}}}(\widehat{\boldsymbol\Omega})=I_\omega^1(\widehat{\boldsymbol{\Omega}})+ I_\omega^2(\widehat{\boldsymbol{\Omega}}) \propto \frac{{\mu_s^{a-1}}\omega^3}{\exp(\hbar\omega/kT_s)-1},
\ee
where $\mu_s$ is the (stellar-frame) cosine between the emission direction ($\widehat{\boldsymbol\Omega}$) and the local magnetic field, and $a\ge 1$ is a beaming parameter (we adopt $a = 1$ for isotropic emission). As discussed above, for $B\lesssim B_{\rm OV}$, 
the X-ray polarization degree is set to $1$ {at lower energies} (i.e., X-mode-dominated) and to $-1$ {at higher energies} (i.e., O-mode-dominated) by comparing $E$ {with} $E_{\rm ad}$, using a smoothed transition (e.g., a {Fermi--Dirac-type} switch).

\subsubsection{General-Relativistic Effects} \label{sec:GR}

To describe photon propagation from the NS surface to the scattering region, general-relativistic effects must be considered. For typical magnetar spin periods $P =2\pi/\Omega = 2$--$12~\mathrm{s}$, rotational frame-dragging is negligible: $\left(R\Omega / c \right)^2 \lesssim 4 \times 10^{-8} R_6^2 P_{1}^{-2}$ \citep{Fernandez2011}. Thus, the Schwarzschild metric suffices.

Key GR effects include gravitational redshift and light bending. As photons travel to $\gtrsim 10R$ where resonant scattering occurs (see below), the metric at the scattering point is almost flat, with $GM/(10Rc^2) \ll 1$. For redshift, since the photon occupation number $\mathcal{N}(\omega) \propto I_{\omega}/\omega^3$ is Lorentz invariant, under the redshift $\omega \to g\omega$ where $g = \big[ 1 - 2GM/(Rc^2) \big]^{1/2}$ the intensity transforms as $I_{\omega} \to g^3 I_{\omega}$. We treat the light-bending effect using the mapping between local emission angle $\alpha$ and asymptotic angle $\psi$ as in \cite{Beloborodov2002}, which is accurate for $R \gtrsim 4GM/c^2$.

\subsection{Resonant Compton Scattering Formalism}

In the outer magnetosphere, photons are scattered by electrons mainly through RCS. For a magnetar with dipole field $B_\mathrm{p}$ and twist parameter $\xi_\tau$ (see {Eq.~}\eqref{ma}), RCS occurs at the radius (neglecting the motion of electrons; see below)
\begin{equation} \label{rsc}
r_{\mathrm{sc}} = 10.5~R \left( \frac{B_{\mathrm{p}}}{10^{14}~\mathrm{G}} \right)^{1/3} \left( \frac{E}{1~\mathrm{keV}} \right)^{-1/3} F(\theta),
\end{equation}
where $F(\theta) = \left(1 - \frac{3}{4} \sin^2\theta\right)^{1/6}$. Given that the Landau level lifetime of gyrating electrons is much shorter than the cyclotron period \citep{Fernandez2011}, electrons can be assumed to remain in the ground state, rendering individual scattering {events} effectively elastic in the electron rest frame (ERF). We neglect non-resonant scattering and recoil in this work.

\subsubsection{Cross Sections}

In the Thomson limit, the non-relativistic RCS cross sections were first derived by {\citet{Ventura1979}} \citep[see also Chapter 4 {of}][]{1992book}. We denote the unit vector along the {propagation} direction {by} $\widehat{\boldsymbol{\Omega}}$, and the polarization states {by} $\alpha \in \{1, 2\} = \{\mathrm{O, X}\}$. In {the} ERF, the differential cross section takes the form \citep[e.g.,][]{Nobili2008a}:
\begin{equation} \label{diffXsection}
\left(\frac{{\mathrm{d}}\sigma}{{\mathrm{d}}\Omega}\right)^{\mathrm{(ERF)}}_{\alpha \widehat{\boldsymbol\Omega}' \to \beta \widehat{\boldsymbol\Omega}} = \left(\frac{e^2}{m_{e}c}\right)~K'_{\alpha\beta}~\delta(\omega - \omega_B),
\end{equation}
where $\omega=E/\hbar$ is the photon frequency, $\omega_B=eB/m_{e}c$ is the electron cyclotron frequency, and the coefficients $K'_{\alpha\beta}$ are
\begin{equation}
\begin{aligned}
& K'_{11} = \frac{3\pi}{8}\left( \widehat{\boldsymbol\Omega}'\boldsymbol\cdot\widehat {\boldsymbol{B}} \right)^2\left( \widehat{\boldsymbol\Omega}\boldsymbol\cdot\widehat {\boldsymbol{B}} \right)^2, \\
& K'_{12} = \frac{3\pi}{8}\left( \widehat{\boldsymbol\Omega}'\boldsymbol\cdot\widehat {\boldsymbol{B}} \right)^2, \\
& K'_{21} = \frac{3\pi}{8}\left( \widehat{\boldsymbol\Omega}\boldsymbol\cdot\widehat {\boldsymbol{B}} \right)^2, \\
& K'_{22} = \frac{3\pi}{8}.
\end{aligned}
\end{equation}
Note that all {coefficients} involving {dot products} of unit vectors should be calculated in {the} ERF, so they are all related to the variable $\beta_e$. When transforming to the stellar frame, relativistic beaming modifies the differential cross section:
\begin{equation}
\left( \frac{{\mathrm{d}}\sigma}{{\mathrm{d}}\Omega} \right)^{\mathrm{(NS)}} = \frac{1}{\gamma_{e}^2 (1-\beta_{e}\mu)^2} \left( \frac{{\mathrm{d}}\sigma}{\mathrm{d}\Omega} \right)^{\mathrm{(ERF)}},
\end{equation}
where $\mu = \widehat{\boldsymbol{\Omega}} \cdot \widehat{\boldsymbol{B}}$ {is measured in the NS frame}, and the delta function must also transform
\begin{equation}
\delta(\omega - \omega_B)^{(\text{ERF})} \to \frac{1}{\gamma_{e}(1 - \beta_{e}\mu)} \delta(\omega - \omega_D)^{(\text{NS})},
\end{equation}
with $\omega_D = \omega_B / [\gamma_{e}(1 - \beta_{e} \mu)]$. We denote the differential cross section in the stellar frame as 
\begin{equation}
\left( \frac{{\mathrm{d}}\sigma}{{\mathrm{d}}\Omega} \right)^{\mathrm{(NS)}}_{\alpha \widehat{\boldsymbol\Omega}' \to \beta \widehat{\boldsymbol\Omega}} = \left(\frac{e^2}{m_{e}c}\right)~K_{\alpha\beta}~\delta(\omega - \omega_D)
\label{NSF}
\end{equation}
with {$K_{\alpha\beta}\equiv K_{\alpha\beta}(\widehat{\boldsymbol{\Omega}}'\to \widehat{\boldsymbol{\Omega}})= K'_{\alpha\beta}/[\gamma_e (1-\beta_e \mu)]^3$} {the non-dimensional scattering coefficients in the NS frame}. The scattering radius for {a} photon of frequency $\omega$ can be {determined} by solving the resonance condition $\omega = \omega_D$ for {a} given emission direction and magnetosphere parameters. For the non-relativistic cases with {a} dipole field, solving $\omega=\omega_D\simeq\omega_B$ yields {Eq.~}\eqref{rsc}.

In the ERF, integrating \eqref{diffXsection} over scattered directions and summing over $\beta$ gives the total cross section $\sigma_{\alpha}(\widehat{\boldsymbol{\Omega}}, \beta_e)$ for photons of mode $\alpha$ propagating along $\widehat{\boldsymbol{\Omega}}$, which is also Lorentz-invariant:
\begin{equation} \label{totXsection}
\begin{aligned}  
\sigma_{1} &= \frac{2\pi^2 e^2 }{m_{e}c}\left(\widehat{\boldsymbol{\Omega}}\boldsymbol \cdot \widehat{\mathbf{B}}\right)^2\delta(\omega - \omega_B),\\
\sigma_{2} &= \frac{2\pi^2 e^2 }{m_{e}c}{\,}\delta(\omega - \omega_B).
\end{aligned}
\end{equation}
Using the estimate $\sigma\simeq (\pi^2 e^2/m_{e}c){\,}\delta(\omega - \omega_B)$ and {Eq.~}\eqref{density} for $n_e$, we obtain the characteristic optical depth across the resonant layer \citep[see also][]{TLK}:
\begin{equation}\label{opticaldepth}
\tau_0(\widehat{\boldsymbol{\Omega}})\equiv \int \mathrm{d}s{\,} n_e\sigma \simeq \frac{\pi\xi_\tau}{4\beta_0}{\,}\frac{B}{r_{\mathrm{sc}}}{\,}\bigg| \frac{\mathrm{d}B}{\mathrm{d}s} \bigg|^{-1} \sim \frac{\pi\xi_\tau}{4\beta_0},
\end{equation}
where $\mathrm{d}s$ is integrated across the RCS layer along $\widehat{\boldsymbol{\Omega}}$ and $\big|{\mathrm{d}B}/{\mathrm{d}s} \big|=(\widehat{\boldsymbol{\Omega}}\boldsymbol{\cdot\nabla})B\sim B/r_{\rm sc}$ is the directional derivative of $B$ along the photon path. We note that the optical depth is independent of energy. For a specific polarization mode $\alpha$, {the corresponding optical depth is
\be\label{eq:opdepth}
\tau_\alpha(\widehat{\boldsymbol{\Omega}})\equiv\int \mathrm{d}s\,\mathrm{d}\beta_e\, n_ef_1(\beta_e){(1-\beta_e\mu)} \sigma_\alpha,
\ee 
where the factor {$(1-\beta_e\mu)$} reflects the encounter rate between photons and electrons due to their relative motion in the NS frame.} Note that $\tau_\alpha\sim 1$ indicates that {the expectation value of the number of scatterings over the photon ensemble is unity} in the magnetosphere.

\subsubsection{Polarized Radiative Transfer in magnetosphere} \label{sec:RTE}

After a photon leaves the NS surface, it {propagates} in the magnetosphere, whose dielectric property is dominated by vacuum polarization \citep{Heyl2003}. At small radii, the photon's polarization state evolves adiabatically following the varying magnetic field it experiences, such that the specific mode intensity is constant (except for the GR corrections discussed in {Section~}\ref{sec:GR})
\footnote{An exception occurs when the photon encounters the quasi-tangential (QT) point, where the photon momentum is nearly aligned with the local magnetic field. The X-ray polarization may change significantly when the photon passes through the QT region \citep{Wang&Lai2009}. Since only a small fraction of the NS surface radiation is affected by the quasi-tangential propagation, we shall neglect this effect in this paper.}.

Near the RCS layer, the transfer equation for polarized intensity $I^\alpha_\omega(\widehat{\boldsymbol\Omega})\equiv I^\alpha_\omega(\widehat{\boldsymbol\Omega}, \boldsymbol{r_\text{sc}})$ (where the $\boldsymbol{r_\text{sc}}$ dependence highlights the dependence on the scattering location) of mode $\alpha$ and frequency $\omega$ is given by (in the NS frame)
\begin{widetext}
\begin{equation}\label{RTE}
\begin{aligned}
\left(\widehat{\boldsymbol{\Omega}}\boldsymbol{\cdot\nabla}\right) I^\alpha_{\omega}({\widehat{\boldsymbol\Omega}})
=& -\int\mathrm{d}\beta_e\, n_e f_1(\beta_e){(1-\beta_e \mu)} \sigma_{\alpha}(\widehat{\boldsymbol{\Omega}}, \beta_e) I^\alpha_{\omega}({\widehat{\boldsymbol\Omega}})
\\
& + \sum_{\beta}  \int\mathrm{d}\beta_e\oint\mathrm{d}\Omega'\, n_e f_1(\beta_e){(1-\beta_e\mu')\,}\eta^{-3} I^\beta_{\eta\omega}(\widehat{\boldsymbol\Omega}') \left( \frac{\mathrm{d}\sigma}{\mathrm{d}\Omega'} \right)^{(\mathrm{NS})}_{\beta \widehat{\boldsymbol\Omega}' \to \alpha \widehat{\boldsymbol\Omega}},
\end{aligned}
\end{equation}
\end{widetext}
{where ${\mu'}=\widehat{\boldsymbol\Omega}'\cdot\widehat{\boldsymbol{B}}$ is measured in the NS frame}, $\eta = \eta(\beta_e, \widehat{\boldsymbol{B}}, \widehat{\boldsymbol\Omega}', \widehat{\boldsymbol\Omega})=(1-\beta_e^2)/[~ (1-\beta_e\cos\theta_1)(1+\beta_e\cos\theta_2') ~]$ is the Doppler-shift factor linking the incident ($\eta\omega$) and scattered ($\omega$) frequencies, $\cos\theta_1=\widehat{\boldsymbol{\Omega}}'\cdot\widehat{\boldsymbol{B}}$ and $\cos\theta_2'=\widehat{\boldsymbol{\Omega}}\cdot\widehat{\boldsymbol{B}}$ are calculated in the NS frame and ERF, respectively, and the $\eta^{-3}$ factor represents the {Doppler} intensity rescaling through scattering (i.e.{,} Comptonization). {Note also that the factors {$(1-\beta_e\mu)$} and {$(1-\beta_e\mu'_0)$} in {Eq.~}\eqref{RTE} reflect the encounter rate due to relative motion. }

We denote the position vector of a small emission patch on the NS surface {by} $R\widehat{\boldsymbol{\Omega}}_s$ and the position vector of the scattering location {by} $\boldsymbol{r_\text{sc}}=r_\text{sc}\widehat{\boldsymbol{\Omega}}_0$. Then the incoming direction vector can be calculated as $\widehat{\boldsymbol\Omega}'=(r_\text{sc}\widehat{\boldsymbol{\Omega}}_0-R\widehat{\boldsymbol\Omega}_s)/|r_\text{sc}\widehat{\boldsymbol{\Omega}}_0-R\widehat{\boldsymbol\Omega}_s|$. The first term on the RHS of {Eq.~}\eqref{RTE} represents the attenuation of the incoming intensity along $\widehat{\boldsymbol\Omega}$, and the second term represents the scattered-in {contribution}. For non-relativistic electrons ($\beta_e \ll 1$), given the narrowness of the resonance, the scattering occurs on a thin surface. For single {scattering}, {Eq.~}\eqref{RTE} can be solved to yield the relationship between the incoming (before RCS) intensity $I_{\omega, \mathrm{in}}^\alpha$ and the outgoing (after RCS) intensity $I_{\omega, \mathrm{out}}^\alpha$ as
\begin{equation}\label{1st_solution_total}
    I_{\omega,\text{out}}^\alpha(\widehat{\boldsymbol{\Omega}})=I_{\omega, \text{out}}^{\alpha\ (\text{I})}(\widehat{\boldsymbol{\Omega}})+I_{\omega, \text{out}}^{\alpha\ (\text{II})}(\widehat{\boldsymbol{\Omega}}).
\end{equation}
The first term is the attenuated incoming intensity that propagates along $\widehat{\boldsymbol{\Omega}}$:
\begin{equation}\label{1st_solution_1}
I_{\omega,\mathrm{out}}^{\alpha\ (\text{I})}(\widehat{\boldsymbol\Omega}) = I_{\omega, \mathrm{in}}^\alpha(\widehat{\boldsymbol\Omega}) \exp\left(-\tau_{\alpha}\right),
\end{equation}
where $\tau_\alpha(\widehat{\boldsymbol{\Omega}})=\int \mathrm{d}s\,\mathrm{d}\beta_e\, n_ef_1(\beta_e){(1-\beta_e\mu)}\sigma_\alpha$ {is the RCS optical depth for mode $\alpha$}. The second term in {Eq.~}\eqref{1st_solution_total} represents the scattered-in part, which is given by
\begin{equation}\label{1st_solution_2-1}
\begin{aligned}
I_{\omega,\mathrm{out}}^{\alpha\ (\text{II})}(\widehat{\boldsymbol\Omega}) = &\sum_\beta \int \mathrm{d}s\,\mathrm{d}\beta_e\oint\mathrm{d}\Omega'\, n_e f_1(\beta_e) \\
& \times {(1-\beta_e\mu')\,}\eta^{-3}I_{\eta\omega,\mathrm{in}}^\beta(\widehat{\boldsymbol{\Omega}}') \left( \frac{\mathrm{d}\sigma}{\mathrm{d}\Omega'} \right)^{\mathrm{(NS)}}_{\beta \widehat{\boldsymbol\Omega}' \to \alpha \widehat{\boldsymbol\Omega}}.
\end{aligned}
\end{equation}
In {Eq.~}\eqref{1st_solution_2-1}, the integration over $\mathrm{d}s$ can be simplified using the delta function in the cross section, and the integration over $\mathrm{d}\Omega'$ for a given scattering site $\boldsymbol{r}_{\text{sc}}$ can be written as
\begin{equation}
    \oint \mathrm{d}\Omega' = \oint \frac{\mathrm{d}\boldsymbol{A}_s\boldsymbol\cdot\widehat{\boldsymbol{\Omega}}'}{\left|\boldsymbol{r}_{\text{sc}}-R\widehat{\boldsymbol{\Omega}}_s\right|^2}=\int\frac{\mathrm{d}A_s \left(\widehat{\boldsymbol{\Omega}}_s\boldsymbol{\cdot\widehat{\Omega}}'\right)}{\left|\boldsymbol{r}_{\text{sc}}-R\widehat{\boldsymbol{\Omega}}_s\right|^2}\Theta\left(\widehat{\boldsymbol{\Omega}}_s\boldsymbol{\cdot\widehat{\Omega}}'\right),
\end{equation}
where $\mathrm{d}\boldsymbol{A}_s=\widehat{\boldsymbol{\Omega}}_s\mathrm{d}A_s$ is the {surface-element} vector of the emission spot on the NS surface (see {Figure~}\ref{fig:schematic}). Note that we have included a Heaviside function $\Theta\left(\widehat{\boldsymbol{\Omega}}_s\boldsymbol{\cdot\widehat{\Omega}}'\right)$ to ensure that only photons emitted outward from the surface are accounted for. Substituting the expression of $n_e$ and $(\mathrm{d}\sigma/\mathrm{d}\Omega')$ with the simplifications above, {Eq.~}\eqref{1st_solution_2-1} can be rewritten as
\begin{widetext}
\begin{equation}\label{1st_solution_2-2}
\begin{aligned}
I_{\omega,\mathrm{out}}^{\alpha\ (\text{II})}(\widehat{\boldsymbol\Omega}) = &
\int\frac{\mathrm{d}A_s\left(\widehat{\boldsymbol{\Omega}}_s\boldsymbol{\cdot\widehat{\Omega}}'\right)\Theta\left(\widehat{\boldsymbol{\Omega}}_s\boldsymbol{\cdot\widehat{\Omega}}'\right)}{4\pi \left|\boldsymbol{r}_{\text{sc}}-R\widehat{\boldsymbol{\Omega}}_s\right|^2} \sum_\beta\int \mathrm{d}\beta_e\ f_1(\beta_e)
K_{\beta\alpha}(\widehat{\boldsymbol{\Omega}}'\to\widehat{\boldsymbol{\Omega}}) \\
& \times{(1-\beta_e\mu')} \eta^{-3} I_{\eta\omega,\mathrm{in}}^\beta(\widehat{\boldsymbol{\Omega}}')\frac{\xi_\tau}{\beta_0} \frac{B}{r_\mathrm{sc}}\left| \frac{\partial}{\partial s}\left[ \frac{B}{\gamma_e(1-\beta_e\mu)} \right] \right|^{-1},
\end{aligned}
\end{equation}
\end{widetext}
where $K_{\beta\alpha}(\widehat{\boldsymbol{\Omega}}'\to\widehat{\boldsymbol{\Omega}})$ is given by {Eq.~}\eqref{NSF}. {Eqs.~}\eqref{1st_solution_total}, \eqref{1st_solution_1}, and~\eqref{1st_solution_2-2} give the total outgoing intensity after one scattering. We expect that the single-scattering approximation is valid when $\tau_\alpha \lesssim 1$. We may still apply it for the $\tau_\alpha \gtrsim 1$ cases to show qualitative physical {trends}. Higher-order solutions can be obtained perturbatively, analogous to {those} in \cite{L&G2006}, where an analytical solution for the one-dimensional case is given.

After RCS, an outgoing photon continues to propagate in the magnetosphere, with its polarization state evolving adiabatically, i.e., $I_\omega^\alpha(\widehat{\boldsymbol\Omega})={\text{constant}}$. Eventually, the photon reaches the ``polarization-limiting radius'' $r_{\rm pl}$, beyond which the polarization state is frozen. The value of $r_{\rm pl}$ depends on the photon energy $E$, the dipole field strength $B_\text{p}${,} and the magnetar rotation rate, and generally $r_{\rm pl}/R\gtrsim 100$ \citep{vanAdelsberg2006}. Since $r_{\rm pl}\gg r_{\rm sc}$, regardless of the structure and shape of the RCS layer, the radiation emerging from the layer with mode intensities $I_\omega^\alpha$ evolves adiabatically in the magnetosphere such that the radiation at $r > r_{\rm pl}$ consists of approximately the same X-mode and O-mode intensities as those emerging from the RCS layer, with a small mixture of circular polarization generated around $r_{\rm pl}$. See \cite{Lai2023} for more details.

\subsection{Computational Workflow}

In our magnetar radiation model, different physical processes are separated cleanly by their characteristic radial scales: emission near the surface, propagation and scattering through the magnetosphere, and polarization evolution from the scattering layer to the polarization-limiting radius. Our calculations are divided into three steps:

\begin{figure*}
\centering
\includegraphics[width=0.95\linewidth]{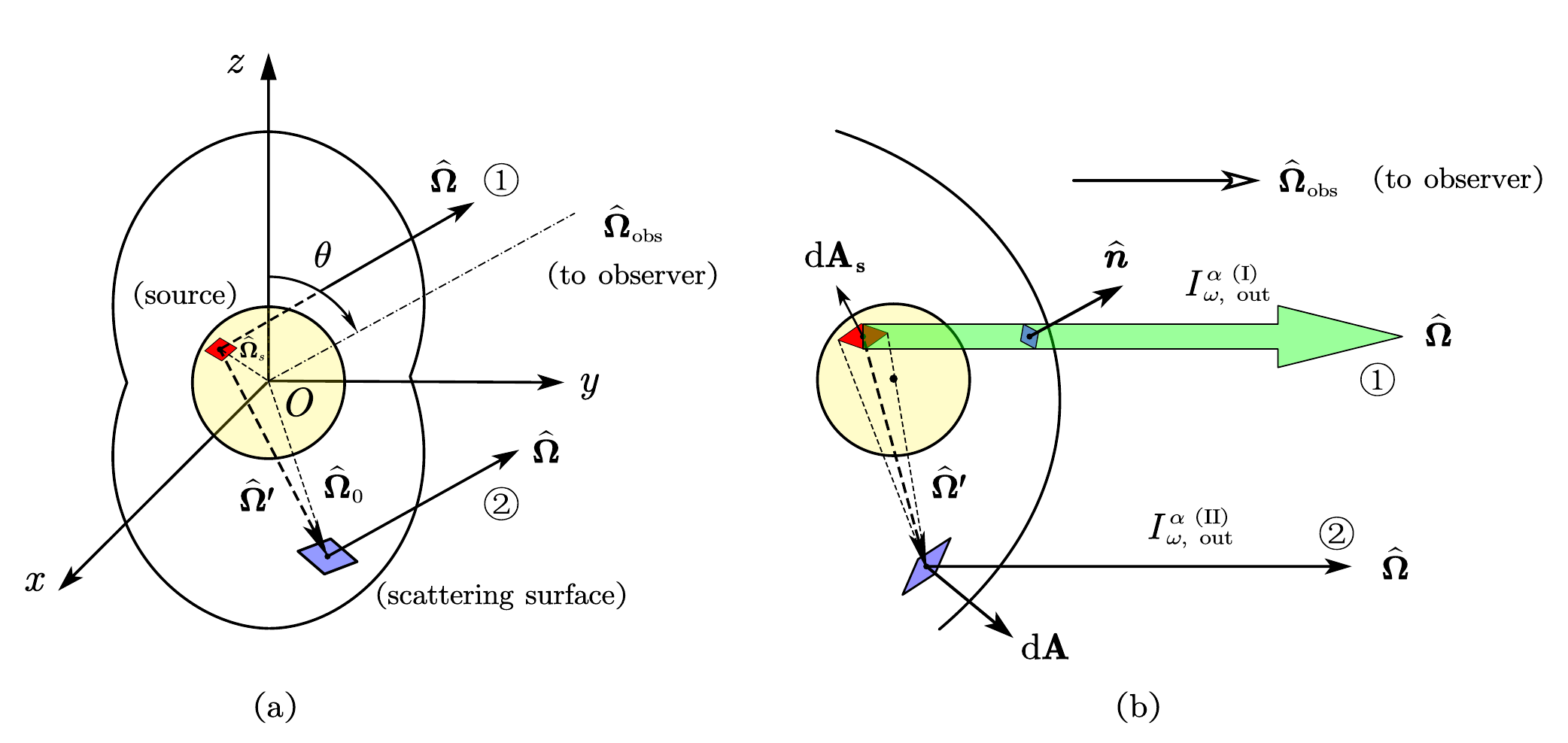}
\caption{Schematic {illustration} of the radiative transfer with resonant Compton scattering and the computation of the observed polarization flux. The surface element of the NS is ${\mathrm{d}}\boldsymbol A_s=\widehat{\boldsymbol\Omega}_s{\mathrm{d}}A_s$. The scattering surface element (located at $\boldsymbol r_{\text{sc}}=r_{\text{sc}}\widehat{\boldsymbol{\Omega}}_{\text{sc}}$) is denoted by ${\mathrm{d}}\boldsymbol A$. The scattered radiation propagates along the direction $\widehat{\boldsymbol{\Omega}}\simeq \widehat{\boldsymbol{\Omega}}_{\text{obs}}$, and the incident radiation (before scattering) propagates along $\widehat{\boldsymbol\Omega}'=(r_\text{sc}\widehat{\boldsymbol{\Omega}}_0-R\widehat{\boldsymbol\Omega}_s)/|r_\text{sc}\widehat{\boldsymbol{\Omega}}_0-R\widehat{\boldsymbol\Omega}_s|$. The magnetic axis of the magnetar is along the $z$-axis, and $\widehat{\boldsymbol{\Omega}}_\text{obs}$ is in the $yz$ plane, with the angle between $\widehat{\boldsymbol{\Omega}}_\text{obs}$ and the $z$-axis being $\theta$.}
\label{fig:schematic}
\end{figure*}

\begin{enumerate}
\item \textit{Emission near the surface under vacuum resonance}: Near the neutron star surface and atmosphere, the dominant process is polarized X-ray emission influenced by QED vacuum resonance. This region has been extensively studied \citep[e.g.,][]{vanAdelsberg2006,Lai2023,Ruth2024}, allowing us to directly adopt previously computed emissivity profiles as input. In this work, we use a simplified prescription for the surface emission, as described in {Section~}\ref{sec:surfaceemission}.

\item \textit{Adiabatic propagation from the surface and RCS}: As discussed, resonant scattering occurs at $\sim 10 R$. After emission from the NS surface, the photon polarization evolves adiabatically following the magnetic field and propagates solely under gravitational effects until scattering. Light bending is handled using the method of \cite{Beloborodov2002}. Specifically, the light-bending mapping between local emission angle $\alpha$ and asymptotic angle $\psi$ corresponds to a mapping between $\left(\widehat{\boldsymbol{\Omega}}_s\boldsymbol{\cdot\widehat{\Omega}}'\right)$ at {the} RCS layer and that at the NS surface, which is used in {Eq.~}\eqref{1st_solution_2-2}. Note that the dot product in {Eq.~}\eqref{1st_solution_2-2} represents the cosine of the asymptotic angle after light bending. We shall see later that this mapping is also applied in calculating the flux contribution of $I_{\omega,\text{out}}^{\alpha \ (\text{I})}$, where a similar dot product also {appears}.  

\item \textit{Continuing adiabatic propagation to large distances where polarization freezes}: After scattering, the polarization continues to evolve adiabatically until the photon reaches the {polarization-limiting} radius ($\gg r_\text{sc}$), where the polarization freezes and aligns coherently, yielding the observable net polarization. The observed flux is computed by integrating the contributions from all locations on the scattering surface. Specifically, we integrate the mode-resolved intensity $I_{\mathrm{out}, \omega}^\alpha \left(\widehat{\boldsymbol{\Omega}}, \boldsymbol{r}_{\mathrm{sc}}(\widehat{\boldsymbol\Omega}_0)\right)$ over the outer surface of the scattering layer.
\end{enumerate}

Figure~\ref{fig:schematic} illustrates the geometric framework for our calculations of the magnetar X-ray radiation. Panel (a) defines the geometry: we set up a Cartesian coordinate system with the magnetic axis along the $z$-axis, and the observer's line of sight (LOS) is in the $yz$ plane at angle $\theta$ from the $z$-axis. The radiation source (position vector $R\widehat{\Omega}_s$) is located at $(\theta_s, \phi_s)$ on the NS surface. The location on the scattering surface is denoted by $\boldsymbol{r}_{\mathrm{sc}}=r_\text{sc}\widehat{\boldsymbol{\Omega}}_0$. The direction of the incoming photon is $\widehat{\boldsymbol\Omega}'$, and that of the scattered photon is $\widehat{\boldsymbol\Omega}$. Two photon populations contribute to the observed flux, as discussed in {Section~}\ref{sec:RTE} and illustrated in {Figure~}\ref{fig:schematic}:
\begin{itemize}
\item \textit{Attenuation term} $I_{\omega, \text{out}}^{\alpha\ (\text{I})}(\widehat{\boldsymbol{\Omega}})$ {---} photons emitted directly toward the observer, attenuated by scattering;
\item \textit{Scattered-in term} $I_{\omega, \text{out}}^{\alpha\ (\text{II})}(\widehat{\boldsymbol{\Omega}})$ {---} photons from all other directions scattered into the LOS.
\end{itemize}

Panel (b) of {Figure~}\ref{fig:schematic} shows how scattered rays reach the observer. Ignoring stellar surface occultation, the total polarized flux in mode $\alpha$ is given by
\begin{equation}\label{flux-total}
\mathcal F_{\omega}^\alpha =\mathcal F_{\omega}^{\alpha\ (\text{I})} + \mathcal F_{\omega}^{\alpha\ (\text{II})} = \int I_{\omega, \text{out}}^{\alpha}(\widehat{\boldsymbol{\Omega}})\left( \widehat{\boldsymbol{\Omega}} \cdot \widehat{\boldsymbol{\Omega}}_{\text{obs}} \right) \mathrm{d}\Omega.
\end{equation}
Since the rays reaching the observer from different locations on the scattering surface are nearly parallel, we have $\left( \widehat{\boldsymbol{\Omega}} \cdot \widehat{\boldsymbol{\Omega}}_{\text{obs}} \right)=1$. Substituting {Eqs.~}\eqref{1st_solution_total}, \eqref{1st_solution_1}, and~\eqref{1st_solution_2-2} into {Eq.~}\eqref{flux-total}, using $\mathrm{d}\Omega = |\mathrm{d}\boldsymbol{A\cdot}\widehat{\boldsymbol{\Omega}}_\text{obs}|/D^2$ with $\mathrm{d}\boldsymbol{A}$ being the {area-element} vector of the scattering surface and $D$ being the distance to the observer, we obtain
\begin{widetext}
\begin{equation}\label{flux-expression}
\begin{aligned}
\mathcal F_{\omega}^{\alpha\ (\text{I})}
  &= \frac{A_s}{D^2}\left(\widehat{\boldsymbol{\Omega}}_s\cdot\widehat{\boldsymbol{\Omega}}_{\text{obs}}\right)
     I_{\omega, \mathrm{in}}^\alpha(\widehat{\boldsymbol{\Omega}}_{\text{obs}})
     \exp(-\tau_{\alpha}),\\
\mathcal F_{\omega}^{\alpha\ (\text{II})}
  &= \frac{A_s}{4\pi D^2}\frac{\xi_\tau}{\beta_0}\sum_\beta
     \int \mathrm{d}\beta_e \oint \mathrm{d}\Omega_0\ f_1(\beta_e)\eta^{-3}
     I_{\eta\omega,\mathrm{in}}^\beta(\widehat{\boldsymbol{\Omega}}') K_{\beta\alpha}(\widehat{\boldsymbol{\Omega}}'\to\widehat{\boldsymbol{\Omega}}_{\text{obs}})\\
  &\quad\times {(1-\beta_e\mu')}\frac{r^2_{\text{sc}}\left| \hat{\boldsymbol{n}}\cdot \widehat{\boldsymbol{\Omega}}_{\text{obs}} \right|}
    {\left| \boldsymbol{r}_{\text{sc}}- R\widehat{\boldsymbol{\Omega}}_s \right|^2}
     \left(\widehat{\boldsymbol{\Omega}}_s\cdot\widehat{\boldsymbol{\Omega}}'\right)
     \Theta\!\left(\widehat{\boldsymbol{\Omega}}_s\cdot\widehat{\boldsymbol{\Omega}}'\right)
     \frac{B}{r_\text{sc}}
     \left| \frac{\partial}{\partial s}\left[ \frac{B}{\gamma_e(1-\beta_e\mu)} \right] \right|^{-1},
\end{aligned}
\end{equation}
\end{widetext}
where we have assumed a small emission surface area $A_s$ on the NS surface --- in general, an integration over the whole {NS} surface can be performed. In {Eq.~}\eqref{flux-expression}, we have used $\mathrm{d}\boldsymbol{A}=\hat{\boldsymbol{n}}\ r_\text{sc}^2\mathrm{d}\Omega_0$, where $\hat{\boldsymbol{n}}$ is the surface normal vector of the scattering surface. Note that the final form of the first expression in {Eq.~}\eqref{flux-expression} does not involve {an} integration over the scattering surface, since only photons emitted directly toward the observer without scattering contribute to this term. Also note that, to simplify {notation}, the dot products $\left(\widehat{\boldsymbol{\Omega}}_s\cdot\widehat{\boldsymbol{\Omega}}_{\text{obs}}\right)$ and $\left(\widehat{\boldsymbol{\Omega}}_s\cdot\widehat{\boldsymbol{\Omega}}'\right)$ in {Eq.~}\eqref{flux-expression} do not include GR {effects}. As mentioned before, when substituting the incoming intensity $I_{\omega, \text{in}}^\alpha$, we must account for GR light bending from the NS surface to the scattering layer. In this context, the dot products $\left(\widehat{\boldsymbol{\Omega}}_s\cdot\widehat{\boldsymbol{\Omega}}_{\text{obs}}\right)$ and $\left(\widehat{\boldsymbol{\Omega}}_s\cdot\widehat{\boldsymbol{\Omega}}'\right)$ represent the effective projection of the emission area onto the incoming direction $\widehat{\boldsymbol{\Omega}}'$ after light bending. Similarly, by {mapping} $\left(\widehat{\boldsymbol{\Omega}}_s\cdot\widehat{\boldsymbol{\Omega}}_{\text{obs}}\right)$ back to the NS surface using the method of \cite{Beloborodov2002}, we can incorporate the GR {light-bending} effect in {Eq.~}\eqref{flux-expression}.

After we obtain the fluxes $\mathcal F^\alpha_\omega$ of each mode, the linear polarization degree can be calculated as
\begin{equation}\label{PL}
    P_{\mathrm{L}}(\omega)=\frac{\mathcal{F}_\omega^2-\mathcal{F}_\omega^1}{\mathcal{F}_\omega^2+\mathcal{F}_\omega^1}.
\end{equation}

\section{Result of Simplified Fiducial Model} \label{sec:validation}

To identify how each ingredient in our semi-analytic treatment shapes the observables, we begin with a simplified model for preliminary analysis. Throughout this section, we adopt a pure dipole field with a cold-plasma approximation, ignoring relativistic effects. The electron number density follows {Eq.~}\eqref{density}, with the parameter $\zeta\equiv\xi_\tau/\beta_0$ controlling the electron density (see {Eq.~}\eqref{density}) and hence the effective RCS strength (see {Eq.~}\eqref{opticaldepth}), independently of the field geometry. Under these assumptions, the mode-resolved optical depths can be calculated as $\tau^{(\text{NR})}_\alpha=\int{\mathrm{d}}s{\,} n_e\sigma_\alpha^{(\text{NR})}$ and the flux can be simplified to
\begin{equation}\label{flux-simplified}
    \begin{aligned}
    \mathcal F_{\omega}^{\alpha (\text{NR,I})}=& \frac{A_s\widehat{\boldsymbol{\Omega}}_s\cdot\widehat{\boldsymbol{\Omega}}_{\text{obs}}}{D^2}I_{\omega, \text{in}}^\alpha(\widehat{\boldsymbol{\Omega}}_\text{obs})\exp\left(-\tau_{\alpha}^{(\text{NR})}\right), \\
    \mathcal F_{\omega}^{\alpha\ (\text{NR,II})} =& \frac{A_s\zeta}{4\pi D^2}\sum_\beta \oint \mathrm{d}\Omega_0\ K_{\beta\alpha} I_{\omega,\mathrm{in}}^\beta(\widehat{\boldsymbol{\Omega}}')~\frac{B}{r_\text{sc}}\left| \frac{{\mathrm{d}}B}{{\mathrm{d}}s} \right|^{-1} \\
    &\times \frac{r^2_{\text{sc}}\left| \hat{\boldsymbol{n}}\boldsymbol\cdot \widehat{\boldsymbol{\Omega}}_{\text{obs}} \right|}{\left| \boldsymbol{r}_{\text{sc}}- R\widehat{\boldsymbol{\Omega}}_s \right|^2} \left(\widehat{\boldsymbol{\Omega}}_s\boldsymbol{\cdot\widehat{\Omega}}'\right)\Theta\left(\widehat{\boldsymbol{\Omega}}_s\boldsymbol{\cdot\widehat{\Omega}}'\right).
    \end{aligned}
\end{equation}

For our fiducial setup, we adopt magnetar parameters $B_{\mathrm p} = 5\times10^{13}{~\mathrm{G}}$, $R = 10{~\mathrm{km}}$, and $kT_s = 0.6{~\mathrm{keV}}$, with the emission point located at $(\theta_s, \phi_s) = (65^\circ, 30^\circ)$ and the observer's viewing angle set to $\theta = 25^\circ$. \added{The magnitude of $B_{\rm p}$ is consistent with the $B\lesssim B_{\rm OV}$ regime in polarized surface emission.} We compute the polarized fluxes and linear polarization degree in the IXPE band ($2$--$8~\mathrm{keV}$).

\begin{figure*}
\centering
\includegraphics[width=\linewidth]{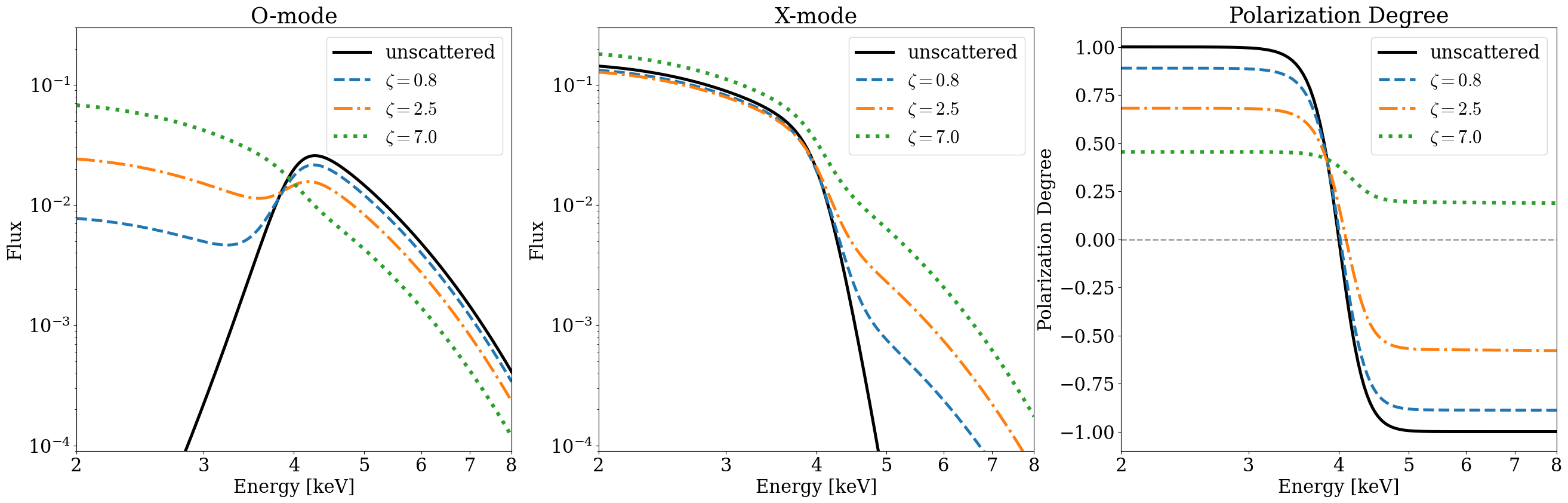}
\caption{Flux and polarization spectra for different scattering strengths compared to the unscattered case in the simplified setup with $B_{\mathrm p} = 5\times10^{13}{~\mathrm{G}}$, $kT = 0.6{~\mathrm{keV}}$, $R = 10{~\mathrm{km}}$, a hot spot at $(\theta_s,\phi_s) = (65^\circ,30^\circ)$, viewing angle $\theta=25^\circ$, and varying $\zeta \equiv \xi_\tau/\beta_0 = 0.8, 2.5, 7.0$ (see {Eqs.~}\eqref{density} and~\eqref{opticaldepth}). \textit{Left and middle}: the relative O-mode and X-mode fluxes. \textit{Right}: the corresponding linear polarization degree $P_L(\omega)$ as a function of energy.}
\label{fig:flux_pol}
\end{figure*}

Figure~\ref{fig:flux_pol} illustrates the impact of varying $\zeta$ on both the mode-resolved spectra and polarization of the observed radiation. For weaker {scattering} with small $\zeta$ (where the single-scattering approximation is valid), the initially dominant mode is attenuated, while the subdominant mode gains {power} through {scattering}, reducing the O/X {spectral} contrast. For stronger {scattering} with larger $\zeta$, the total flux increases, since the scattered-in term scales directly with electron density. As the density becomes too high, our single-scattering approximation may overestimate the flux. In this regime, the spectral shapes of the O- and X-modes become similar.

The right panel of Figure~\ref{fig:flux_pol} shows the corresponding linear polarization degree $P_{\rm L}(\omega)$ as a function of photon energy. As the RCS strength increases, $P_{\mathrm{L}}$ flattens, and it becomes positive for all photon energies in the case of $\zeta=7$. Thus, the original mode-switching signature may disappear in strong-scattering cases. As we will see later, incorporating {special-relativistic} effects may erase the mode-switching signature {more efficiently}, even for cases with low optical depths. This phenomenon will be discussed in more detail in {Sections~}\ref{sec:results} and~\ref{sec:conclusion}.

Regarding the validity of our calculations, all the results above rely on the single-scattering approximation, i.e., a small resonant optical depth. Our numerical checks confirm that the estimated optical depth ($\sim 0.1$--$1$) from {Section~}\ref{sec:model} is valid. Specifically, for the geometry used above, we find $\tau \lesssim 1$ when $\zeta \lesssim 3.7$, supporting the accuracy of the single-scattering solution in this range. Notably, the disappearance of the {mode switch} in our calculations occurs at $\zeta \gtrsim 5.6$, where the single-scattering approximation becomes questionable.

\added{For the high magnetospheric-density cases, the scattered-in term is proportional to $n_e$ and may therefore be overestimated in the first-order treatment. However, a fully self-consistent post-scattering optical depth cannot be assigned within the simplified cold, non-relativistic setup used in this section. In this baseline model the resonance is represented by a geometrically thin surface, and the velocity distribution is not resolved. However, any real current-conducting plasma has a finite velocity width such that a scattered photon can still resonate with electrons in the remaining part of the velocity distribution, and a photon scattered back toward the star can encounter additional resonant surfaces \citep[another example of self-consistent treatment can be seen in][]{Hukun2026}. A physical value of $\tau_{\alpha,\rm esc}$ therefore requires a genuine multiple-scattering calculation with a specified velocity distribution, scattering geometry, and photon trajectory.\footnote{The discussion on the contribution from higher-order terms in the following paragraphs aims at clarifying the reliability of the result that strong RCS may erase the mode-switching feature in the non-relativistic regime; we do not include the effect of $\tau_{\text{esc}}$ in Section~\ref{sec:results}}}

\added{Here we use two limiting prescriptions only to place a limit on the possible impact of post-scattering attenuation. The unweighted scattered-in term corresponds to our first-order baseline calculation, while the deliberately conservative prescription $\tau_{\alpha,\rm esc}=\tau_\alpha$ gives an upper-bound estimate of the suppression of the first-scattered component. In the latter case we define, with $\tau_\alpha$ given by Eq.\eqref{eq:opdepth},}
\begin{equation}\label{flux-simplified-limit}
    \begin{aligned}
    \mathcal F_{\omega, \rm limit}^{\alpha ({\rm NR,I})}
    &= \mathcal F_{\omega}^{\alpha ({\rm NR,I})}, \\
    \mathcal F_{\omega, \rm limit}^{\alpha ({\rm NR,II})}
    &= \mathcal F_{\omega}^{\alpha ({\rm NR,II})}\exp(-\tau_\alpha).
    \end{aligned}
\end{equation}

{{\begin{figure*}
\centering
\includegraphics[width=\linewidth]{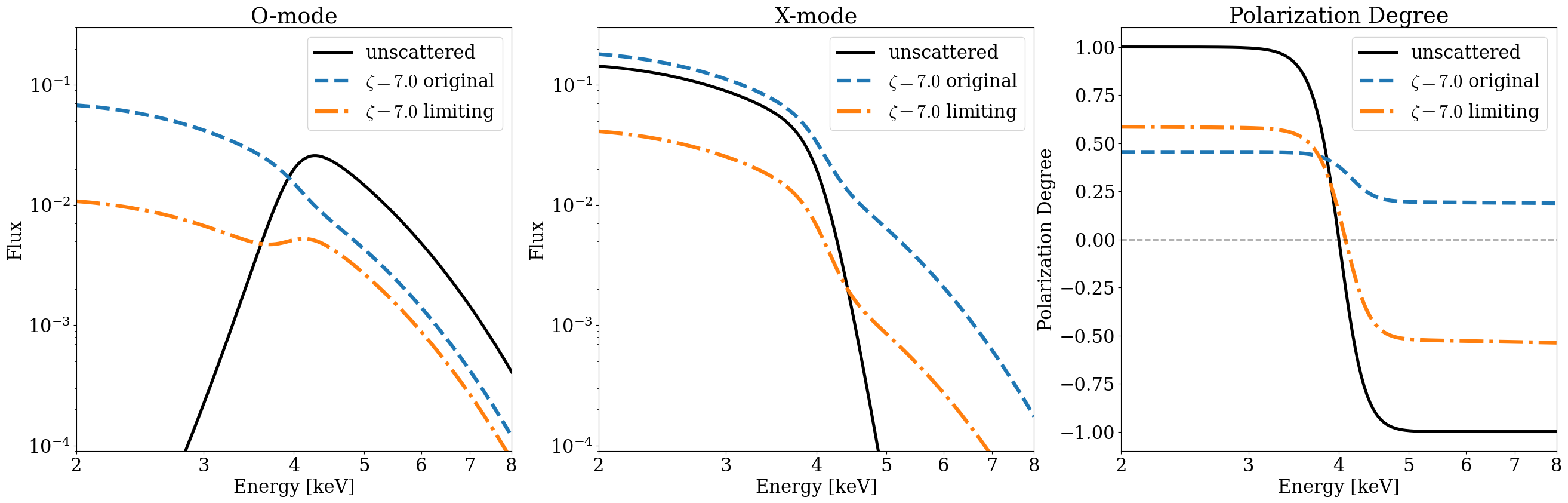}
\caption{\added{The relative O-mode and X-mode fluxes and polarization spectra for two first-order prescriptions (see Eq.~\eqref{flux-simplified-limit}) for the scattered-in component, compared with the unscattered case in the simplified model setup. The blue curves show the unweighted first-scattered baseline, while the orange curves apply the $\exp(-\tau_\alpha)$ attenuation factor to the scattered-in component. Other parameters are $B_{\mathrm p} = 5\times10^{13}~\mathrm{G}$, $kT = 0.6~\mathrm{keV}$, $R = 10~\mathrm{km}$, a hot spot at $(\theta_s,\phi_s) = (65^\circ,30^\circ)$, viewing angle $\theta=25^\circ$, and $\zeta \equiv \xi_\tau/\beta_0 = 7.0$ (see Eqs.~\eqref{density} and~\eqref{opticaldepth}).}}
\label{fig:flux_pol-3}
\end{figure*}}

Figure~\ref{fig:flux_pol-3} shows the results for $\zeta=7.0$, where $\tau_{\alpha, \text{esc}}=0$ and $\tau_\alpha$ are adopted respectively as two limiting cases. As shown in the figure, after introducing a larger $\tau_{\alpha, \text{esc}}$, the flux is substantially suppressed and therefore does not violate energy conservation. However, because the scattered-in contribution is also significantly weakened, the behavior originally indicated in the right panel, namely that RCS may erase the mode-switching feature, becomes less evident. 

Note that in our full-model study in Section \ref{sec:results}, once relativistic effects are included, RCS can still robustly produce the disappearance of the mode-switching behavior in the cases that generally satisfy $\tau\lesssim 0.4$. Thus, despite the single-scattering approximation, our calculation provides useful guidance on the effects of RCS on the magnetar soft X-ray polarization spectra.}

\section{Full Model: Parameter Study} \label{sec:results}

In this section, we adopt the full model setup, including both the thermal velocity distribution of magnetospheric electrons and {(both special and general)} relativistic effects. When the electron velocities are thermal about a mean drift, the resonance condition $\omega = \omega_D(\boldsymbol{r}, \mu, \beta_e)$ no longer selects a single geometric surface but broadens into a thin region centered around the drift-defined resonant surface, with thickness set by the velocity distribution.

To include the GR effects, we fix the compactness of the NS at
\begin{equation}
u \equiv \frac{2GM}{Rc^2} = 0.3,
\end{equation}
and the other NS parameters are held consistent with the previous section. Since all photons experience nearly the same redshift from emission to scattering in our geometries, GR mainly introduces a common rescaling in the intensity and energy. Therefore, we interpret the surface temperature $kT_s$ as the effective blackbody temperature of X-ray emission just before scattering, the same as in Section~\ref{sec:validation}.

Under the current model, the scattering is controlled by the parameter set:
\begin{equation}
\{ \xi_\tau, \beta_0, kT_e, kT_s, \theta, M \}.
\end{equation}
In the survey below, we fix $kT_s$, $E_{\mathrm{ad}}${,} and $M$, and vary $(\xi_\tau, \beta_0, kT_e, \theta)$, while all other parameters are the same as in {Section~}\ref{sec:validation}.

\subsection{Optical-depth Dependence}\label{sec:validity}

Before analyzing the spectra and polarization, it is useful to delineate under what conditions the single-scattering solution is valid. The resonant optical depth $\tau_\alpha$ (for mode $\alpha$) is primarily controlled by $(\xi_\tau, \beta_0)$, which set the electron density and relativistic beaming.

\begin{figure}
\centering
\includegraphics[width=\linewidth]{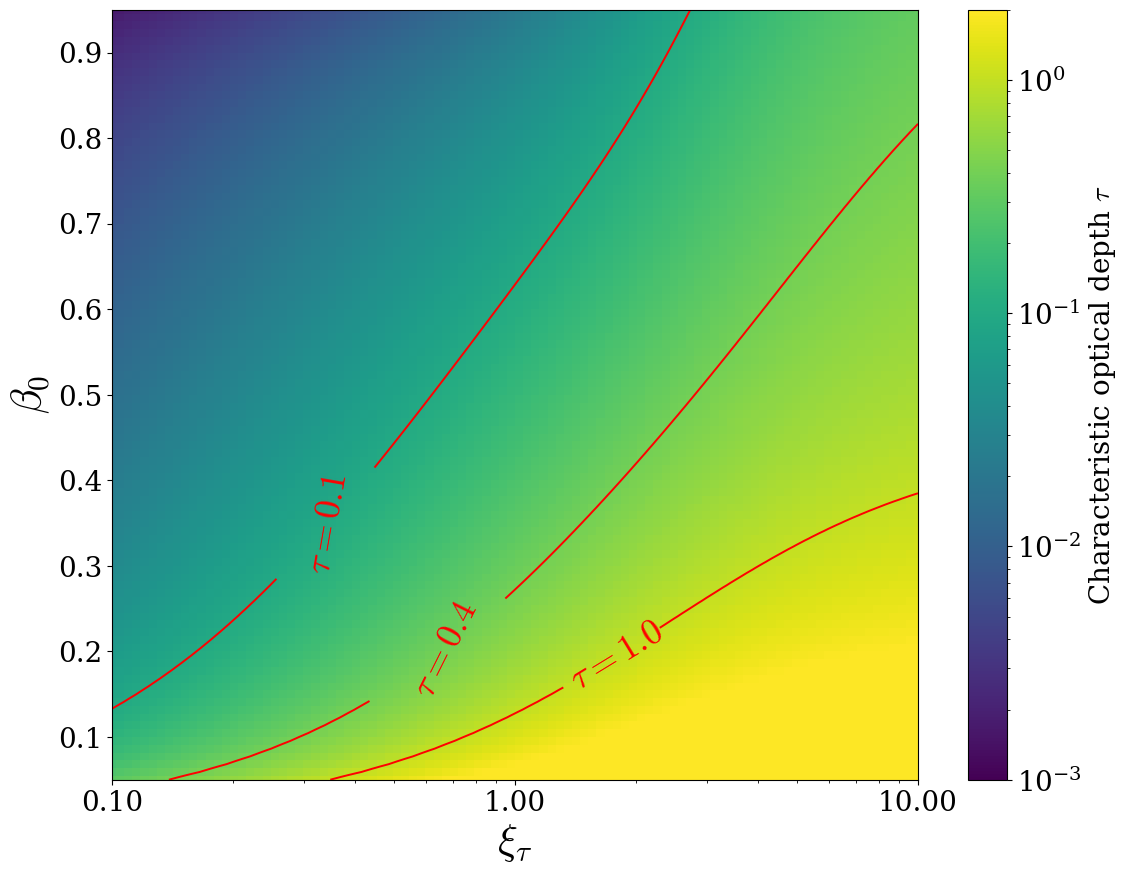}
\caption{Heatmap of the characteristic resonant optical depth in the full model. The parameters are: $B_\mathrm p = 5\times10^{13}~\mathrm{G}$, $R = 10~\mathrm{km}$, $kT_e = 10~\mathrm{keV}$, $kT_s = 0.6~\mathrm{keV}$, a hot spot at $(\theta_s, \phi_s) = (53^\circ, 37^\circ)$, and viewing angle $\theta=35^\circ$. Contours for $\tau_0 = 1.0$, $\tau_0 = 0.4$ and $\tau_0=0.1$ delineate the validity {regime} of the single-scattering/first-order approximation.}
\label{fig:opticaldepth}
\end{figure}

Figure~\ref{fig:opticaldepth} presents the heatmap of the characteristic optical depth $\tau_0$ (see {Eq.~}\eqref{opticaldepth}), calculated at $kT_e = 10~\mathrm{keV}$ and $\theta = 35^\circ$, as functions of $\xi_\tau$ and $\beta_0$. The $\tau_0=1$ contour marks the single-scattering boundary; the $\tau_0=0.4$ contour corresponds to $\lesssim 10\%$ error for a first-order expansion of $\exp(-\tau)$. As expected, larger $\xi_\tau$ and smaller $\beta_0$ yield higher $\tau_\alpha$ due to the increased particle density. {Physically, the optical-depth map delineates the regime of applicability for the results in this section: for parameter combinations with $\tau\lesssim 1$, the single-scattering approximation may be regarded as approximately valid; for parameter combinations with $\tau\lesssim 0.4$, the single-scattering solution is expected to be reasonably reliable. }

{In addition, the optical-depth map also provides a useful reference for interpreting the results in the subsequent parameter study: for instance, if a certain behavior is observed in the regime $\tau\lesssim 0.4$, it can be confidently attributed to the RCS effect; if it is observed in the regime $\tau\gtrsim 1$, it should not be regarded as a strict quantitative prediction of RCS. }

Using the twist parametrization defined in {Eq.~}\eqref{ma}, the total pole-to-pole twist angle is approximately
\begin{equation}
\Delta\phi = \int_0^\pi \frac{B_\phi}{B_\theta \sin\theta}~{\mathrm{d}}\theta \sim \pi \xi_\tau,
\end{equation}
so the ranges adopted in previous Monte Carlo studies (e.g., \citealt{Nobili2008a}, \citealt{Fernandez2011}) fall largely within the $\tau_0 \lesssim 1$ regime in Figure~\ref{fig:opticaldepth}.

\subsection{Spectra and Polarization} \label{sec:spectra}

We {survey} four key parameters: the viewing angle $\theta$, the magnetic twist $\xi_\tau$, the mean drift speed of electrons $\beta_0$, and the plasma temperature $kT_e$. The observed flux is computed as a {multidimensional} integral over the emission direction, scattering geometry, and electron velocity. Given the smoothness of the integration kernel, we evaluate the integral via Monte Carlo quadrature with $10^6$ samples per configuration. This choice introduces small statistical fluctuations, or ``noise'', in the results, but enables significantly improved computational speed and numerical stability, resulting in a relative numerical uncertainty below $0.5\%$.

\begin{figure*}
\centering
\includegraphics[width=\linewidth]{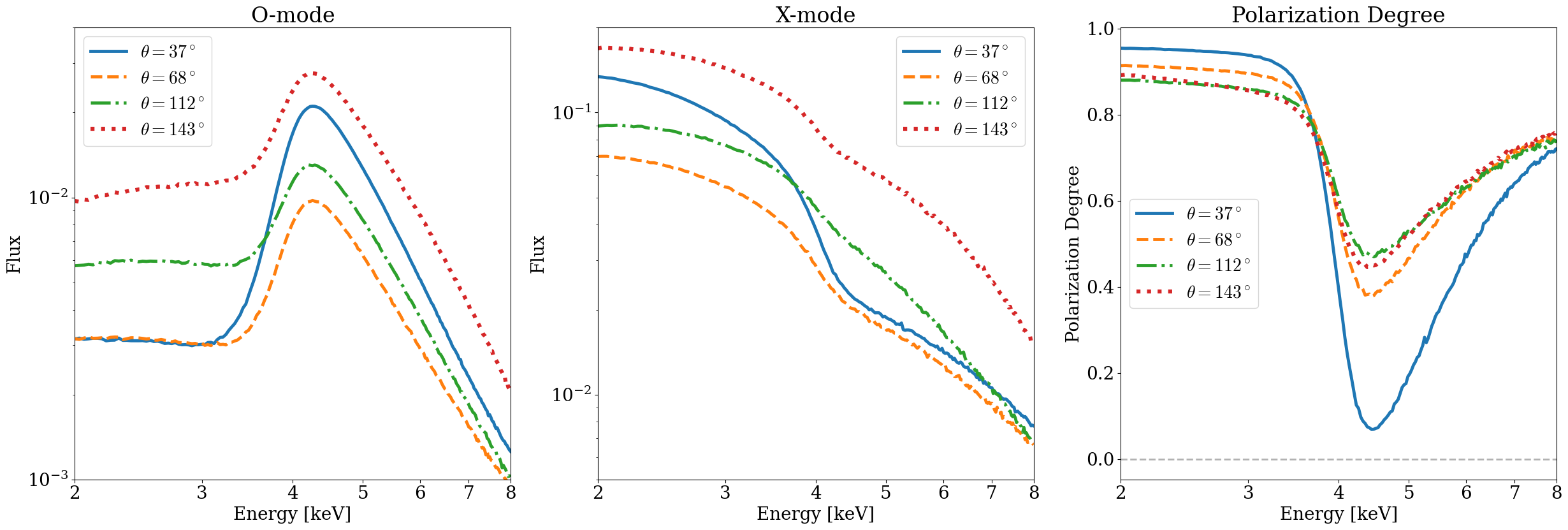}
\caption{
Flux and polarization spectra for different observer viewing {angles} in the full model. The parameters are: $B_\mathrm p = 5\times10^{13}~\mathrm{G}$, $R = 10~\mathrm{km}$, $kT_e = 10~\mathrm{keV}$, $kT_s = 0.6~\mathrm{keV}$, $\xi_\tau = 0.2$, $\beta_0 = 0.5$, and a hot spot at $(\theta_s, \phi_s) = (90^\circ, 37^\circ)$. Results are shown for the viewing angles $\theta = 37^\circ, 68^\circ, 112^\circ, 143^\circ$.
}
\label{fig:viewingangle}
\end{figure*}

\paragraph{Viewing geometry.} Figure~\ref{fig:viewingangle} presents the flux and polarization spectra for various viewing angles. For this plot, we consider a single hot spot located at the equator $(\theta_s, \phi_s) = (90^\circ, 37^\circ)$ to highlight the {north--south} asymmetries \citep[see][]{Nobili2008a}. Notably, pairs of LOS angles symmetric about the equator (e.g., $37^\circ$ and $143^\circ$) yield markedly different flux and polarization spectra, demonstrating the impact of Doppler shifts caused by large-scale net currents flowing from magnetic north to south. Importantly, the polarization curves retain the qualitative behavior previously seen in {Section~}\ref{sec:validation}: RCS may suppress or erase the intrinsic mode-conversion feature that would otherwise appear in the emission. For instance, the curve corresponding to $\theta = 37^\circ$ shows sign flips in polarization degree, whereas for other angles $P_{\mathrm{L}}$ remains positive across the energy band. We verify that in this parameter setting, the optical depth $\tau_\alpha(\theta)$ remains well below unity (typically $\lesssim {0.15}$), indicating that even single-scattering {events} can significantly modify the polarization signature, despite the overall low scattering probability.

\begin{figure*}
\centering
\includegraphics[width=\linewidth]{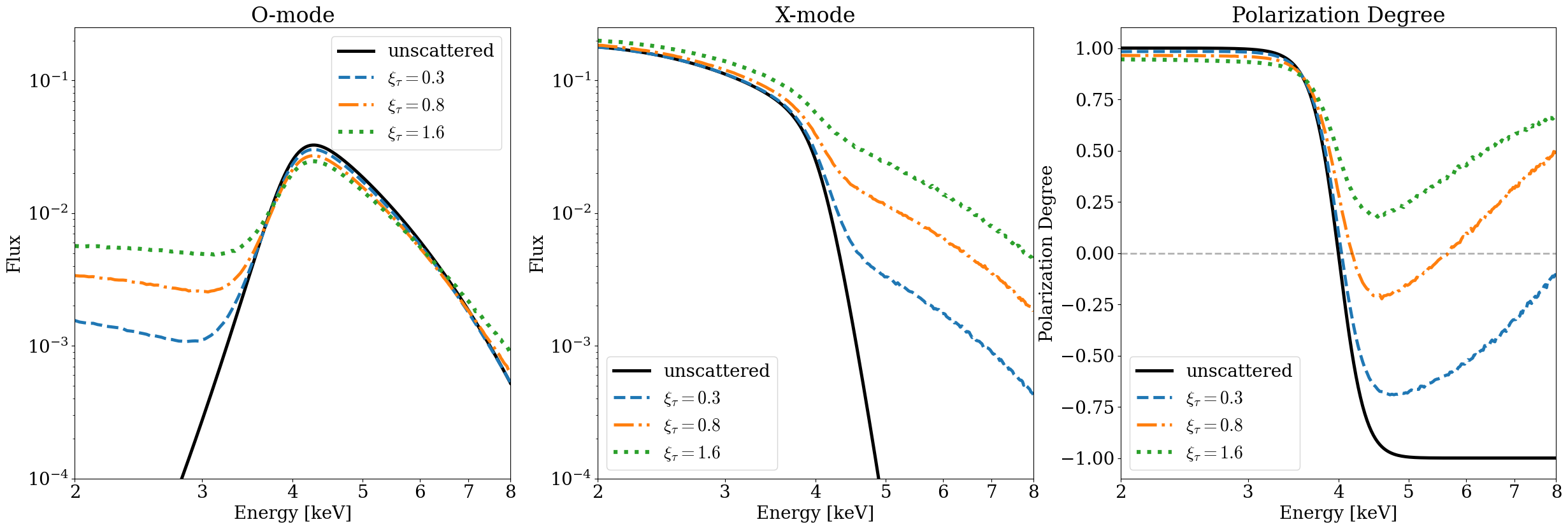}
\caption{
Flux and polarization spectra for different magnetospheric twist parameters in the full model. The parameters are: $B_\mathrm p = 5\times10^{13}~\mathrm{G}$, $R = 10~\mathrm{km}$, $kT_e = 10~\mathrm{keV}$, $kT_s = 0.6~\mathrm{keV}$, $\theta=35^\circ$, $\beta_0 = 0.5$, and a hot spot at $(\theta_s, \phi_s) = (53^\circ, 37^\circ)$. Results are shown for magnetic twist $\xi_\tau = 0.3, 0.8, 1.6$, and are compared with the unscattered case.
}
\label{fig:twist}
\end{figure*}

\paragraph{Magnetic twist.} Figure~\ref{fig:twist} explores how varying the magnetic twist $\xi_\tau$ affects the resulting flux and polarization spectra. The twist primarily modulates the plasma density and hence the scattering strength. As $\xi_\tau$ increases, the flux spectra show progressively larger deviations from the unscattered case, exhibiting the ``spectral smoothing'' noted in {Section~}\ref{sec:validation}. However, relativistic effects become more prominent in the polarization spectra, especially at higher photon energies. Compared to the non-relativistic case, special-relativistic effects cause the polarization degree to rise again at higher energies. In some cases, the polarization curve crosses zero twice within the IXPE band, indicating the {possibility} of multiple $90^\circ$ PA swings (this is also seen in a {pure} QED scenario; see \citealt{Lai2023} and \citealt{Ruth2024}). For sufficiently large $\xi_\tau$, polarization remains strictly positive, and mode switching disappears.

\begin{figure*}
\centering
\includegraphics[width=\linewidth]{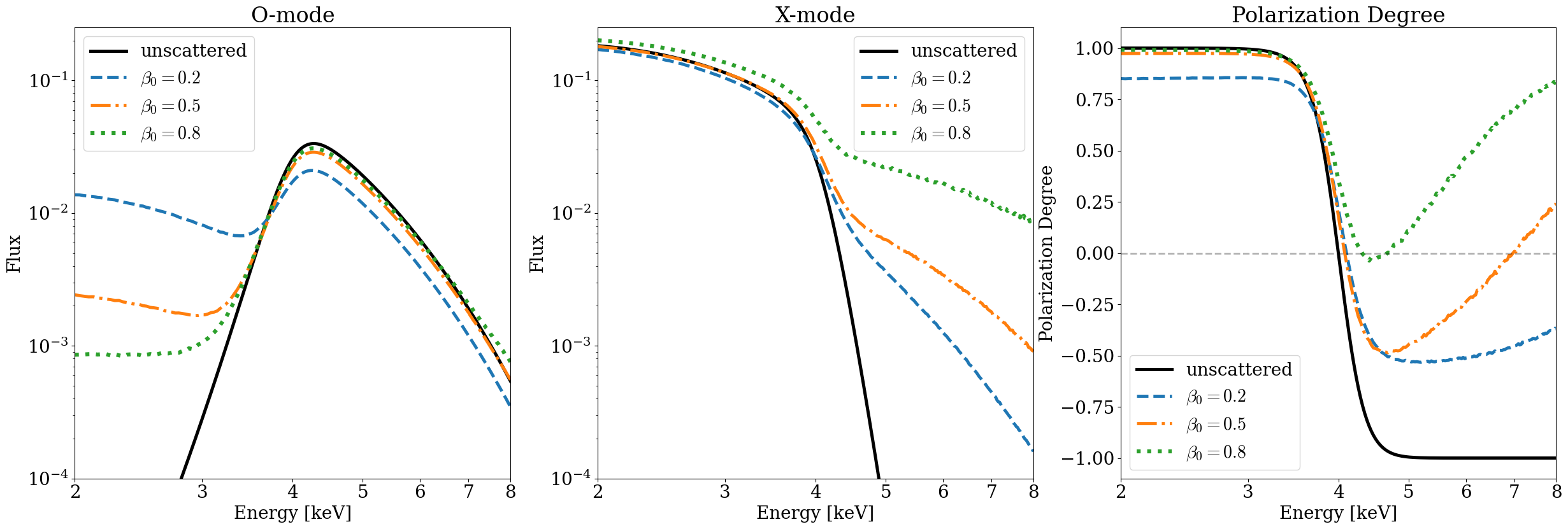}
\caption{
Flux and polarization spectra for different mean electron drift velocities in the full model. The parameters are: $B_\mathrm p = 5\times10^{13}~\mathrm{G}$, $R = 10~\mathrm{km}$, $kT_e = 10~\mathrm{keV}$, $kT_s = 0.6~\mathrm{keV}$, $\theta=35^\circ$, $\xi_\tau = 0.5$, and a hot spot at $(\theta_s, \phi_s) = (53^\circ, 37^\circ)$. Results are shown for $\beta_0 = 0.2, 0.5, 0.8$, and are compared with the unscattered case.
}
\label{fig:drift}
\end{figure*}

\paragraph{Electron drift velocity.} Figure~\ref{fig:drift} shows the impact of changing the average drift velocity $\beta_0$ on the flux and polarization spectra. Like $\xi_\tau$, the parameter $\beta_0$ influences both the plasma density and the strength of SR beaming. While its effects on the flux spectrum resemble those of increasing $\xi_\tau$, the behavior of the polarization curves offers further diagnostic insight. Decreasing $\beta_0$ increases the particle density and weakens SR beaming, suppressing $P_{\mathrm{L}}$ at low energies and flattening its high-energy tail that was initially raised by SR; this is consistent with the non-relativistic trend discussed in {Section~}\ref{sec:validation}. However, in terms of the first zero-crossing energy{---}the location where the initial $90^\circ$ PA swing occurs{---}the effect of varying $\beta_0$ is subtler than that of $\xi_\tau$. The transition slopes remain similar, indicating that $\beta_0$ mainly sets the vertical offset (baseline) of $P_{\rm L}$, while it plays a secondary role in determining the slope that shapes the crossing location.
\footnote{A dynamic analogy may be helpful: imagine {the} polarization curve as a catenary suspended between a high left platform and a vertical wall on the right. Increasing the relativistic effect raises the right anchor, steepening the tail of the curve while leaving its middle largely unchanged.}

\begin{figure*}
\centering
\includegraphics[width=\linewidth]{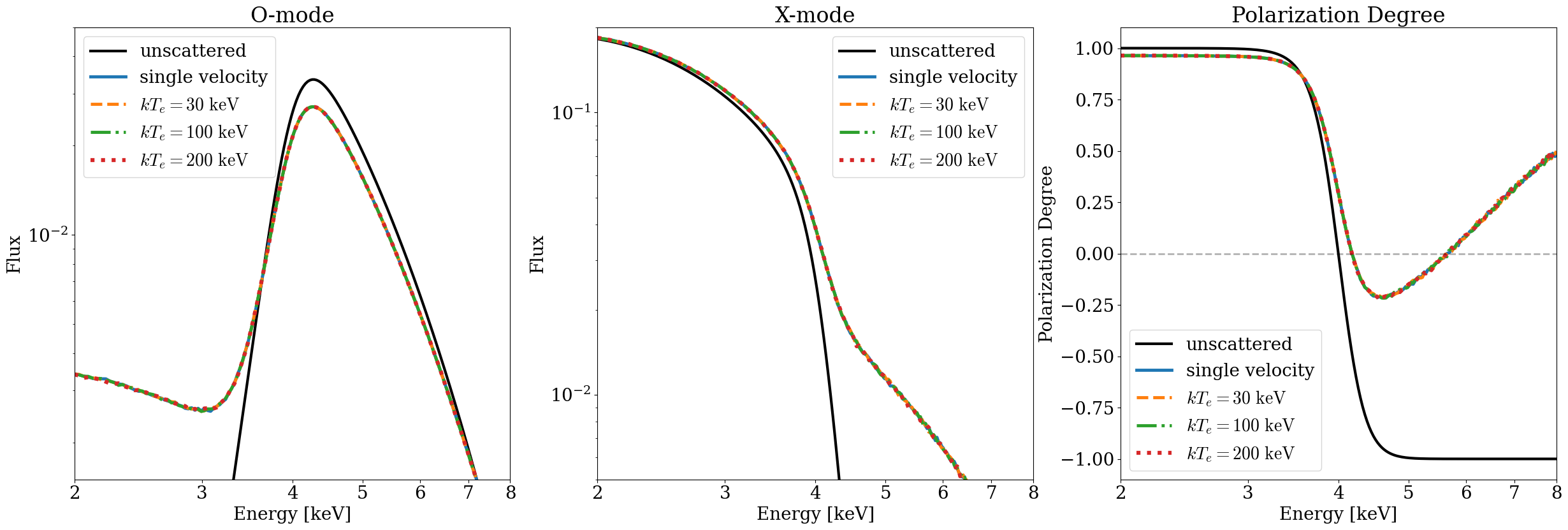}
\caption{
Flux and polarization spectra for different magnetospheric plasma temperatures in the full model. The parameters are: $B_\mathrm{p} = 5\times10^{13}~\mathrm{G}$, $R = 10~\mathrm{km}$, $kT_s = 0.6~\mathrm{keV}$, $\theta=35^\circ$, $\xi_\tau=0.8$, $\beta_0 = 0.5$, and a hot spot at $(\theta_s, \phi_s) = (53^\circ, 37^\circ)$. Results are shown for $kT_e = 30, 100, 200~\mathrm{keV}$, and are compared with the unscattered and single-velocity cases.
}
\label{fig:temperature}
\end{figure*}

Together, these two main control parameters, $\xi_\tau$ and $\beta_0$, determine how RCS reshapes polarization spectra mainly through {changing} magnetospheric density and {special-relativistic} effects. Two principal effects emerge: (i) a reduction in both the absolute value and spectral variation of the polarization degree; and (ii) a high-energy rise of $P_{\rm L}$ driven by SR mode redistribution. Even in the single-scattering regime, RCS may erase the {mode-switch} feature that is originally present in the emission. To understand this trend, consider the polarized RCS cross sections in the ERF (see {Eq.~}\eqref{diffXsection}):
\begin{equation}\label{Xsection}
\begin{aligned}  
\sigma_{1-1} &= \frac{1}{3}\sigma_{1-2} = \frac{\pi^2 e^2 }{2m_{e}c}\delta(\omega - \omega_B)\left(\boldsymbol{\Omega}'\boldsymbol \cdot \widehat{\mathbf{B}}\right)^2,\\
\sigma_{2-2} &= 3\sigma_{2-1} = \frac{3\pi^2 e^2 }{2m_{e}c}\delta(\omega - \omega_B).
\end{aligned}
\end{equation}
These expressions represent the relative probabilities for photons to scatter between specific initial and final polarization states. Notably, regardless of the initial state, scattering tends to populate the X-mode (the polarization index $\alpha=2$) more efficiently. Since the original emission is X-dominated at low energies, RCS has {a} limited effect there, whereas at higher energies (O-dominated) RCS transfers power to the {X-mode}, producing the characteristic high-energy rise of $P_{\rm L}$. The electron density mainly rescales $P_{\mathrm{L}}$, while the SR effects modulate its slope, particularly in the high-energy tail. Together, they reshape the polarization distribution and influence the energy at which mode switches {occur}.

\paragraph{Thermal broadening.} Finally, Figure~\ref{fig:temperature} examines the role of plasma temperature $kT_e$ in shaping the flux and polarization spectra. For visibility, the flux panels are vertically rescaled. Compared to the single-velocity case, thermal motion {induces only} modest changes. Unlike $\beta_0$, which parameterizes the bulk drift, increasing $kT_e$ only broadens the velocity distribution around $\beta_0$. This tends to average out the variation in Doppler effects over electron speeds, resulting in {behavior} that closely resembles the single-velocity case at the same $\beta_0$. This is consistent with previous findings by \citet{Nobili2008a}.

\added{\paragraph{Power-law-like spectral component.}
Previous Monte Carlo simulations of magnetar spectra show that repeated RCS can convert seed thermal photons into a non-thermal component \citep[e.g.,][]{Nobili2008a,Nobili2008b}. However, our single-scattering calculation is not originally designed to produce such a fully developed continuum, but it does capture the onset of spectral hardening by one resonant scattering. To quantify this effect, we fitted the calculated photon spectrum in the IXPE band with the blackbody-plus-power-law model,
\begin{equation}
\begin{aligned}
\dot N(E)=&\dot N_{\rm BB}(E)+\dot N_{\rm PL}(E)\\
    =&A_{\rm BB}{E^2\over \exp(E/kT_s)-1} +A_{\rm PL}E^{-\Gamma_{\rm eff}} ,
\end{aligned}
\end{equation}
where $\dot N $ represents the total photon number flux, and $\dot N_{\rm{BB, PL}}$ are the contributions from the blackbody and power law component, respectively. For the representative $\xi_\tau=0.8$ case in Fig.~\ref{fig:twist}, the XSPEC fit gives $\Gamma_{\rm eff}\simeq 3.0$. The power law-like residual contributes only about $18\%$ of the integrated flux at $\sim 2$ keV and becomes dominant mainly above $\sim 6~{\rm keV}$. A local estimate of the residual slope, $-\mathrm{d}\ln\dot N_{\rm PL}/\mathrm{d}\ln E$, varies strongly across the band from order unity to higher than $4$, rather than forming a stable broad-band power law. We therefore interpret this component as a local hard excess produced by the onset of Comptonization, not as a self-consistent nonthermal continuum. Such a continuum requires higher-order, iterative resonant inverse Compton scatterings. We also note that the blackbody-plus-power-law decomposition is phenomenological; for some other soft-X-ray magnetar spectra, a two-temperature blackbody model can provide a comparably good fit. \citep[e.g.,][]{2026fireball}.}

\section{Summary and Discussion} \label{sec:conclusion}

{In this work,} we have developed a unified and analytically tractable framework to approximate how resonant Compton scattering in the magnetosphere and QED vacuum resonance in the atmosphere jointly shape the X-ray spectra and polarization observed from magnetars. Starting from the polarized radiative transfer equation \eqref{RTE} with scattering, we adopt a first-order approximation in optical depth (Eqs.\eqref{1st_solution_total}, \eqref{1st_solution_1}, \eqref{1st_solution_2-2}), which effectively captures single scattering {events}, to compute the observed polarization fluxes ({Eqs.}\eqref{flux-total}--\eqref{PL}). For a concrete baseline, we adopt an idealized initial surface emission polarization pattern that {mimics} vacuum-resonant conversion---$100\%$ X-mode below $4{~\mathrm{keV}}$ and $100\%$ O-mode above, with a smooth transition in between (\cite{Lai2023})---and then quantify how RCS in the magnetosphere reshapes the observable polarization signatures. Our key findings are as follows:
\begin{enumerate}
    \item Increasing the plasma density in the magnetosphere leads to an overall suppression of both the absolute value and spectral variation of polarization degree $P_{\mathrm{L}}$. In other words, RCS acts to compress the polarization curve toward a horizontal line located \textit{above} zero, diminishing the polarization contrast (see {Figs.~}\ref{fig:flux_pol} and~\ref{fig:twist}).

    \item Increasing the average electron drift velocity enhances the SR effects, which tend to raise $P_{\mathrm{L}}$ in the high-energy tail and steepen its slope. In some cases, an extra mode switch (i.e., sign change in $P_{\mathrm{L}}$, implying a $90^\circ$ PA swing) induced by RCS may occur (see {Figure~}\ref{fig:drift}).

    \item The energy of the \textit{first} mode switch is only weakly shifted by RCS in the parameter ranges explored here; the zero-crossing location remains largely tied to the stellar surface physics set by vacuum resonance.

    \item In the strong RCS regime, the combined effects of increased plasma density and enhanced SR effects may entirely erase the signature of a mode switch within the observed bandpass ($2$--$8~\mathrm{keV}$), even when the QED-induced conversion is present in the initial emission. This is consistent with the mode-redistribution trend expected from the resonant cross sections (see {Eq.~}\eqref{Xsection}).
\end{enumerate}

However, these conclusions should be interpreted in light of several simplifying assumptions. 

(i) Our magnetospheric configuration is \textit{prescribed} rather than dynamically self-consistent. As discussed in {Section~}\ref{sec:model}, the field and current systems are parameterized and do not enforce full force-free equilibrium. In addition, the actual magnetosphere is likely to be {partly} twisted rather than {globally} twisted, and the prescribed thermal velocity distribution in the magnetosphere may not capture some intrinsic physical effects \citep[e.g.,][{who} considered a power-law distribution of relativistic momentum]{Fernandez2007}. {Also, the electron density prescription applied in this work might underestimate the RCS optical depth of the magnetosphere. Indeed, recent kinetic modeling of magnetar coronae suggests that  the pair multiplicities can be moderately high \citep{Thompson2008,Beloborodov2013,Thompson2020,Zhang2025}, i.e., the total density $n_{e^+}+n_{e^-}\gtrsim 30 n_e$. If so, the single-scattering approximation may be questionable. Nevertheless, we still expect that the main trends identified in this work, especially the trend of RCS erasing the mode-switching feature, should be qualitatively robust, since it is primarily driven by the relative scattering probabilities between modes ({Eq.}~\eqref{Xsection}), which are independent of the overall density.}

(ii) Our sample results consider emission from a single surface patch. Radiation from the whole NS surface can be easily included in our formalism and will likely average out some geometric contrasts. \added{Also, a realistic configuration of the extended hot spot may introduce additional behavior in the polarization spectra, which is beyond the scope of this work.}

(iii) Our transport solution is only of first order in optical depth. {Fully} capturing the energy-dependent behavior (e.g., the power-law high-energy tail and Comptonization) and energy/polarization redistribution at high RCS optical depth requires higher-order solutions via iterative scattering treatments \citep[e.g., for 1D iterative {scattering}, see][]{L&G2006}. 

(iv) Our adopted surface polarization pattern is idealized; a more realistic treatment would require detailed modeling of the NS surface and atmosphere.

\added{(v) Finally, we emphasize the restricted domain of validity of the imposed surface-polarization prescription. The atmospheric vacuum-resonance origin of the seed mode switch is self-consistent only when the local magnetic field in the emitting region satisfies $B_{\rm emission}\lesssim B_{\rm OV}$, with $B_{\rm OV}$ given by Eq.~\eqref{eq:Bov}. For the fiducial parameters in our model, this corresponds to $B_{\rm OV}$ of a few $10^{13}~{\rm G}$ across the IXPE band, with variations by factors of a few depending on composition and geometry. This is consistent with the observation of 4U 0142+61 \citep{Sci2022}. If $B_{\rm em}\gtrsim B_{\rm OV}$, the surface emission is expected to remain X-mode dominated over the band, and the atmospheric vacuum resonance alone would not generate the intrinsic $90^\circ$ PA swing assumed in our illustrative seed model. Our results should therefore be interpreted as applying either to low-local-field emitting regions, or to cases where a similar seed polarization pattern is produced by another surface-emission mechanism. The global magnetic-energy budget may still be dominated by hidden internal or toroidal fields, which are not modeled in the present calculation.}

Despite these limitations, the semi-analytic nature of our method offers a major computational advantage and physical transparency. It allows us to extract the main physical trends without the computational cost or complexity of full Monte Carlo simulations and reveals intuitive connections between model inputs and observable outcomes.

Looking ahead, two immediate extensions are natural. First, to better capture viewing-angle-dependent effects, full-surface integration should be implemented, which would remove projection biases from small-area sources and allow more realistic modeling of observation profiles. In our formalism, this only requires treating the coordinates of the surface emission point as integration variables over solid angle. Second, including stellar rotation would allow phase-resolved predictions when the magnetic and spin axes are misaligned. In our framework, this is equivalent to a rotation of the LOS direction vector and tracking how the polarization signal evolves with phase, enabling the calculation of phase-resolved light curves and phase-averaged polarization profiles. These developments, together with higher-order {scattering}, will help connect magnetar surface and magnetosphere models with observations. Upcoming X-ray polarization {missions}, such as eXTP \citep{eXTP2016,eXTP2019}, will have greatly enhanced polarization sensitivity and an extended energy band, providing simultaneous spectral, timing, and polarization data, thereby enriching the magnetar polarization sample for further studies.


\bibliography{ref}{}
\bibliographystyle{aasjournalv7}


\end{document}